# Deriving ventilation imaging from 4DCT by deep convolutional neural network

Yuncheng Zhong[1], Yevgeniy Vinogradskiy[2], Liyuan Chen[1], Nick Myziuk[3], Richard Castillo[4], Edward Castillo[3], Thomas Guerrero[3], Steve Jiang[1], and Jing Wang[1*]

[1]Department of Radiation Oncology, University of Texas Southwestern Medical Center
[2]Department of Radiation Oncology, University of Colorado Denver, Aurora, CO
[3]Department of Radiation Oncology, Beaumont Health System, Royal Oak, MI
[4]Department of Radiation Oncology, Emory University, Atlanta, GA
*Corresponding Author: jing.wang@utsouthwester.edu

## Abstract

**Purpose:** Functional imaging is emerging as an important tool for lung cancer treatment planning and evaluation. Compared with traditional methods such as nuclear medicine ventilation-perfusion (VQ), positron emission tomography (PET), single photon emission computer tomography (SPECT), or magnetic resonance imaging (MRI), which use contrast agents to form 2D or 3D functional images, ventilation imaging obtained from 4DCT lung images is convenient and cost-effective because of its availability during radiation treatment planning. Current methods of obtaining ventilation images from 4DCT lung images involve deformable image registration (DIR) and a density (HU) change-based algorithm (DIR/HU); therefore the resulting ventilation images are sensitive to the selection of DIR algorithms. **Methods:** We propose a deep convolutional neural network (CNN)-based method to derive the ventilation images from 4DCT directly without explicit DIR, thereby improving consistency and accuracy of ventilation images. A total of 82 sets of 4DCT and ventilation images from patients with lung cancer were studied using this method. **Results:** The predicted images were comparable to the label images of the test data. The similarity index and correlation coefficient averaged over the ten-fold cross validation were 0.883±0.034 and 0.878±0.028, respectively. **Conclusions:** The results demonstrate that deep CNN can generate ventilation imaging from 4DCT without explicit deformable image registration, reducing the associated uncertainty.

**Keywords**: lung functional imaging, 4D-CT lung ventilation imaging, convolutional neural network

## I. Introduction

Functional imaging is emerging as a potentially important tool in the planning and evaluation of lung cancer radiotherapy plans. Lung function avoidance has been applied in intensity-modulated radiation therapy (IMRT) planning to limit dose delivered to well-functioning lung parts [1, 2]. Conventional methods to obtain lung ventilation images include nuclear medicine ventilation-perfusion (VQ) scans, positron emission tomography (PET), single photon emission computer tomography (SPECT), or hyperpolarized He-3 MRI [3-5]. Despite providing functional images, these methods suffer some shortcomings including requiring contrast agent administration. In addition, other factors such as availability and cost also limit their applications. Therefore, a more convenient method for lung functional imaging is needed in wider clinical applications.

A 4DCT-based method has been developed to obtain ventilation images. As 4DCT images are easily accessible during radiation treatment planning, 4DCT based ventilation images can be obtained at virtually no additional monetary or dosimetric cost. The method has been validated and evaluated for its ability to generate functional images compared with conventional methods with promising results [5-11]. Lung ventilation images from 4DCT images are calculated according to either Hounsfield unit (HU) changes or regional expansion and contraction of lung volumes. In the method involving HU changes, the temporally resolved reconstructed images at distinct phases within the respiratory cycle are registered, and the HU values are compared based on a simple model of lung tissue as a linear combination of air and 'tissue' components to infer the respiratory volume changes and ventilation images. In the method involving regional expansion and contraction, ventilation images are obtained using the Jacobian of the deformable image registration (DIR) displacement field [7], without explicit accounting of the underlying registered CT image values. Hybrid approaches have also been proposed, whereby the Jacobian determinant is regionally scaled by the ratio of registered HU [12].

The aforementioned approaches rely on the DIR between images at different respiratory phases. Because selecting DIR algorithms affects the stability of the process [13, 14], the accuracy of 4DCT-derived ventilation images is sensitive to the choice of the DIR algorithms [15, 16]. An improved method to derive ventilation images from 4DCT is desired to improve the robustness of the 4DCT-ventilation calculation process.

Deep learning methods have been successfully applied to the field of medical imaging, improving image processing [17], registration [18, 19], de-noising [20, 21], and classification [22]. The deep convolutional neural network (CNN) method extracts features from images and generates model specific parameters [23-30]. CNN can reveal the latent features that are not easily discovered by conventional image processing methods. Successful applications of the deep CNN method on image registration and motion tracking make it a promising approach for ventilation imaging generation from 4DCT. In this work, we propose a deep CNN based method to derive ventilation images directly from the 4DCT without explicit image registration to overcome the uncertainty associated with selecting an accurate DIR algorithm. The proposed method uses 4DCT images as input and the known ventilation images as labeled data for training. Upon training completion, we evaluated the results by quantitatively comparing the predicted results with the reference ventilation images.

## II. Methods and materials

### IIA. Data and pre-processing

We used a supervised machine learning scheme that needs input data and label data to train the models. The input data in our study are 4DCT reconstructed images at the end of inhalation and exhalation stages. The output as the label data are the ventilation images. The workflow and process are illustrated in Fig. 1. In this preliminary study, ventilation images obtained using the DIR and HU change method were used as the labelled data for training [9]. Data for this study was generated as part of prospective functional avoidance clinical trial (NCT02528942). A total of 82 sets of 4DCT and ventilation images from lung cancer patients were used for the study. The majority (72%) of patients had stage III disease and 44% of patients had accompanying chronic obstructive pulmonary disease. Images were pre-processed and randomly divided into training and testing groups.

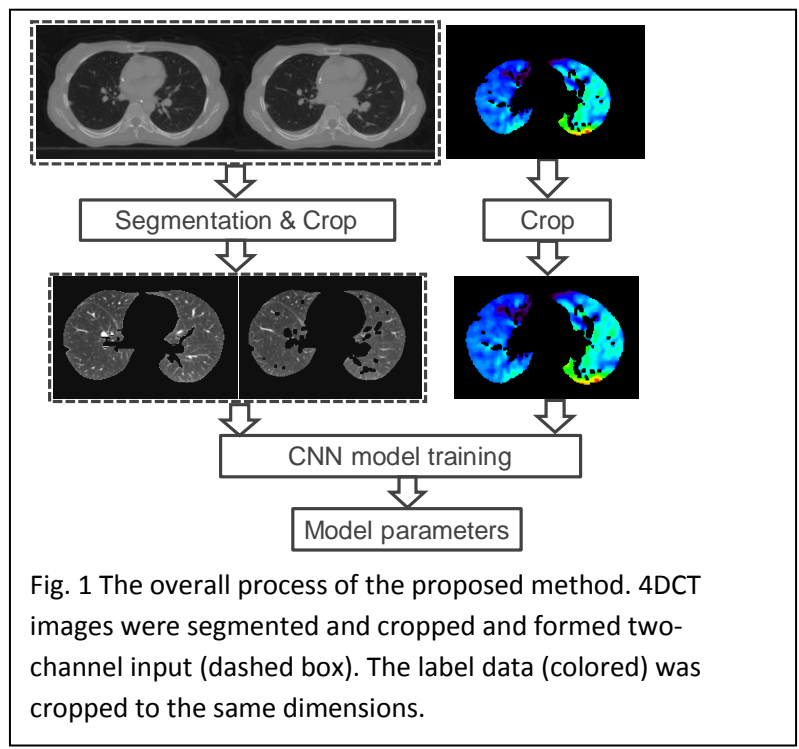

Fig. 1 The overall process of the proposed method. 4DCT images were segmented and cropped and formed two-channel input (dashed box). The label data (colored) was cropped to the same dimensions.

The lungs were segmented on the 4DCT images at the end of inspiration and the end of expiration phases. The segmented images were further cropped to include only the lungs in order to reduce the input image size and computation burden during the model training. Images were segmented using the NIH-CIDI Lung Segmentation tool, which was based on fuzzy connectedness segmentation using graph cut [31]. The segmented lung images were adjusted manually as needed to eliminate other structures, such as heart.

**IIB. Convolutional neural network architecture**

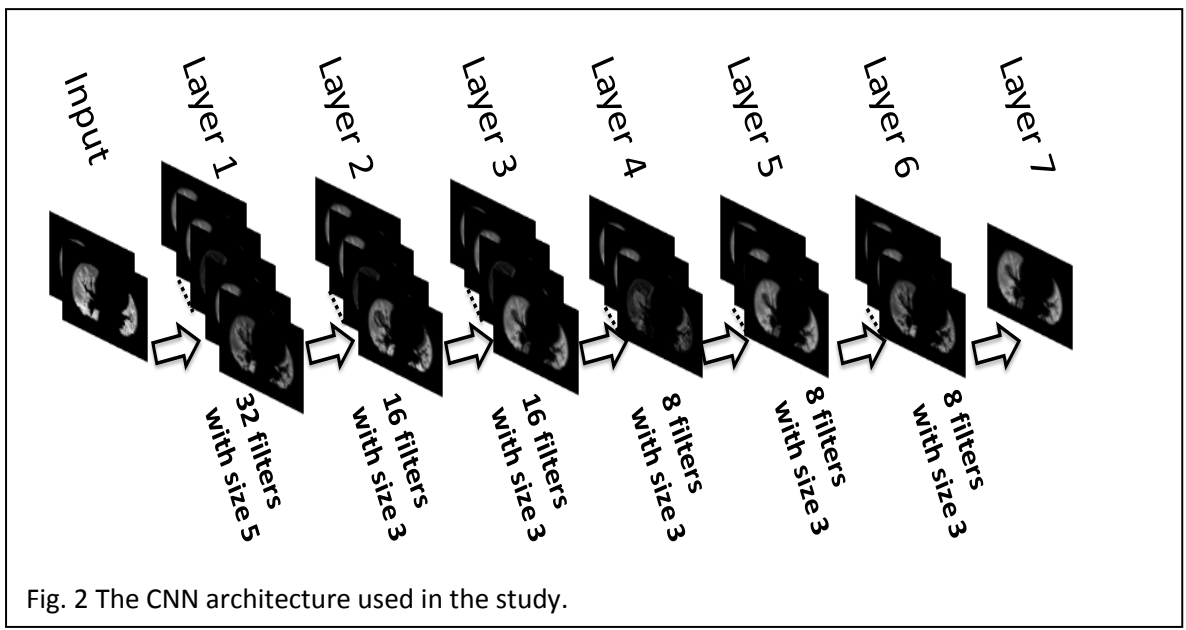

Fig. 2 The CNN architecture used in the study.

The CNN architecture for generating ventilation images based on 4DCT images is shown in Fig. 2. The proposed architecture is composed of one input layer, six convolution layers and one final prediction layer. Images at the end of inhalation and exhalation are utilized as input. In each convolutional layer, a set of learnable small spatially kernels are included to capture local features for each pixel in the images. Zero padding is used in each convolution layer to preserve the output size of each feature map. Additionally, each convolution layer in the proposed architecture is followed by an activation layer that is controlled by an activation function. The activation function that we used in this work is rectifier linear units (ReLU). ReLU has been shown to achieve more efficient gradient propagation and more efficient computation than the logistic sigmoid and hyperbolic tangent activation function. The final prediction layer is also constructed by a convolution layer with kernels of size 3×3×3 and the output is the predicted ventilation image.

To train the proposed CNN model, several key steps are described as follows:

1) Initialization: Initialization of the network weights will affect the convergence of the network. For the proposed CNN model, we used the Xavier initialization method [32], which assigns the network weights from a Gaussian distribution with zero mean and a variance of 1/N where N specifies the number of input neurons. Through Xavier initialization, the variance of input and output for each layer is the same to avoid gradients vanishing or explosion in back-propagation and to ensure activation functions can work normally.

2) Loss function: The loss function that we used to train the proposed CNN model is mean-squared-error (MSE) between the predicted image and label image.

3) Optimization algorithm: We used Adam optimization algorithm which works well in practice and compares favorably to other stochastic optimization methods, to train the proposed CNN model in our experiments. It computes individual adaptive learning rates for different parameters from estimates of first and second moments of the gradients. The initial learning rate and maximum epoch for training in this work are set as $8\times10^{-5}$ and 50, respectively.

We conducted ten-fold cross validation of the patient data to validate model optimization. Data were randomly divided into 10 groups with 8 or 9 patients within a group. Datasets from 9 groups were assembled for model training and the remaining group was used to test the predicted model. During the training process, we use 90 percent of training data to train the model and 10 percent as validation data to select the best model. We then used this selected model to evaluate its performance on the testing data. After each group has been used as the testing data, we obtained 10 sets of results. The average quantities over these 10 sets were then calculated.

**IIC. Evaluation**

Upon completion of training, the model was applied to the independent test dataset with 8 or 9 patients, depending on the random selection from the 10 groups. The predicted ventilation images were compared with the reference images generated with the standard DIR-based algorithm. We used the structure similarity index (SSIM) and correlation coefficient (Spearman's r) as metrics to evaluate the comparison of the predicted and labeled images [33, 34]. The structure similarity is described as

$$SSIM(x,y) = [l(x,y)]^\alpha \cdot [c(x,y)]^\beta \cdot [s(x,y)]^\gamma \qquad (1)$$

with

$$l(x,y) = \frac{2\mu_x\mu_y + C_1}{\mu_x^2 + \mu_y^2 + C_1}$$

$$c(x,y) = \frac{2\sigma_x\sigma_y + C_2}{\sigma_x^2 + \sigma_y^2 + C_2}$$

$$s(x,y) = \frac{\sigma_{xy} + C_3}{\sigma_x\sigma_y + C_3}$$

where $l(x,y), c(x,y), s(x,y)$ represent comparison of luminance, contrast, and structure, respectively; $\mu_x, \mu_y$, $\sigma_x, \sigma_y$ and $\sigma_{xy}$ represent mean, standard deviation, and cross-covariance, respectively; $C_1$, $C_2$ represent the square of 1% and 3% of the dynamic range value, respectively; $C_3 = C_2/2$; $\alpha, \beta, \gamma$ represent the weighting for each comparison, respectively.

The correlation coefficients were computed according to the following equation:

$$\rho(x,y) = \frac{1}{N-1}\sum_{i-1}^{N}\left(\frac{x_i - \mu_x}{\sigma_x}\right)\left(\frac{y_i - \mu_y}{\sigma_y}\right) \qquad (2)$$

where $\mu_x, \mu_y$, $\sigma_x, \sigma_y$ are defined above; N indicates the number of samples; $x_i$, $y_i$ indicate the samples.

## III. Results

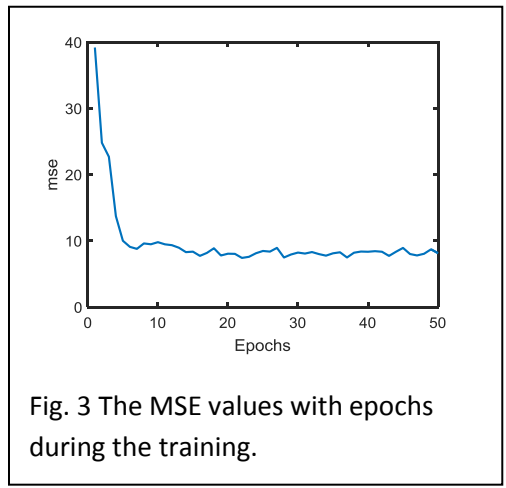

Fig. 3 The MSE values with epochs during the training.

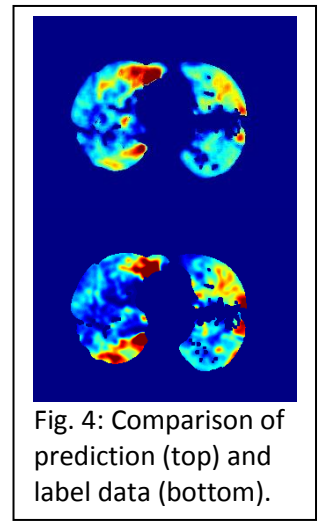

Fig. 4: Comparison of prediction (top) and label data (bottom).

The MSE values converges with epochs in the training (Fig. 3). The training stops after 50 epochs to ensure completion of the minimization process. The visual similarity between the predicted and the label images is shown in Fig. 4 for a representative axial slice (patient CU26B). These images demonstrate good correspondence in their morphology and intensity distribution. The comparison of other testing data shows similar results (Fig. 5). In general, the predicted images (Fig. 5 c) of the testing data are similar to those of the label data (Fig. 5 d) in terms of shape and intensity distribution with local deviations. The overlay of the ventilation images with the chest images provides an overview of the lung ventilation images (Fig. 5e).

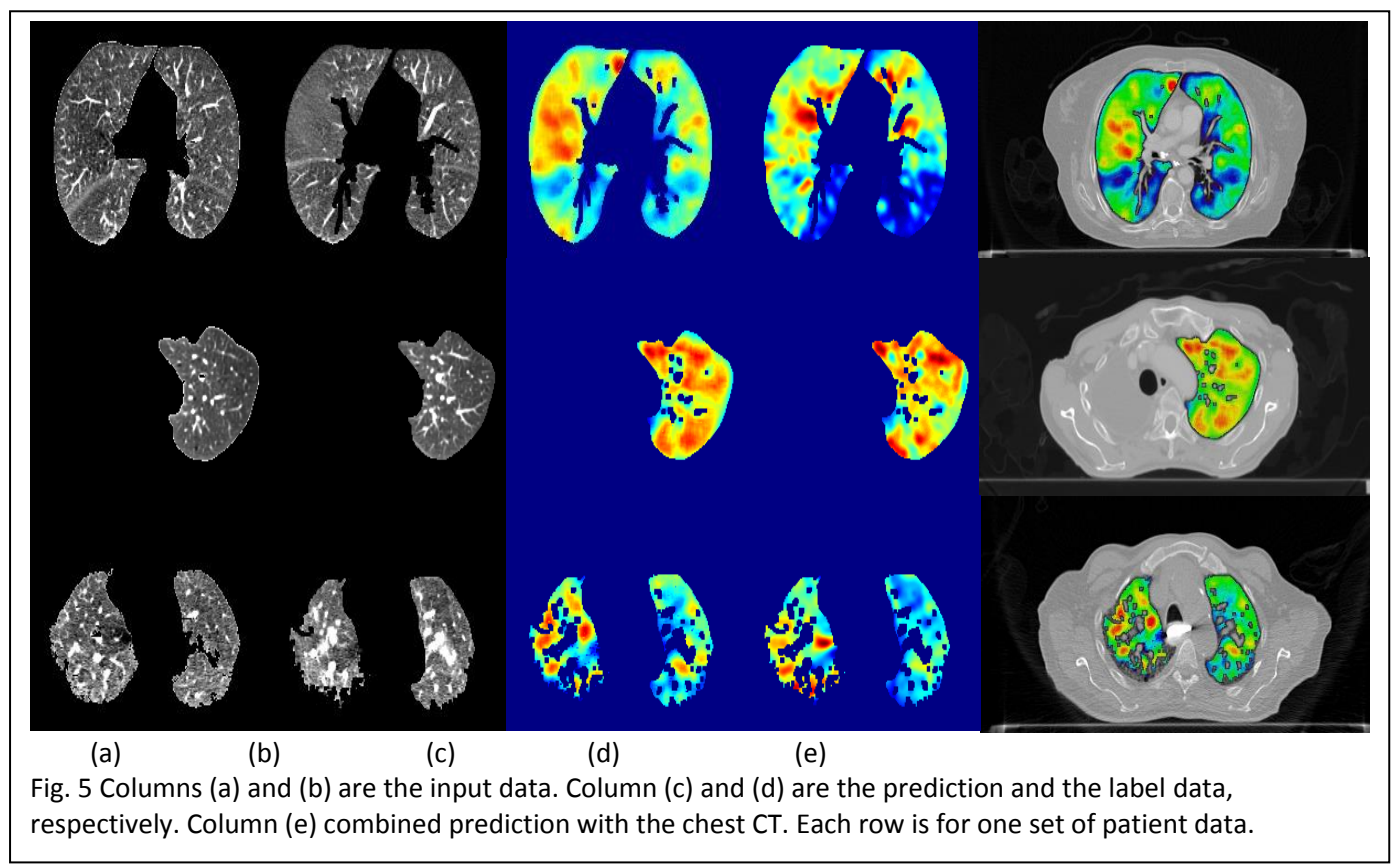

(a)        (b)        (c)        (d)        (e)

Fig. 5 Columns (a) and (b) are the input data. Column (c) and (d) are the prediction and the label data, respectively. Column (e) combined prediction with the chest CT. Each row is for one set of patient data.

SSIM and correlation coefficients calculated from 10-fold cross validations were listed in Tables I and II. The overall average values were 0.883±0.034 and 0.878±0.028 for SSIM and correlation coefficients, respectively.

Table I: SSIM values between predicted and label images for 10-fold cross validation

| Case | 1 | 2 | 3 | 4 | 5 | 6 | 7 | 8 | Average±S.D. |
|---|---|---|---|---|---|---|---|---|---|
| f1 | 0.973 | 0.942 | 0.948 | 0.895 | 0.959 | 0.948 | 0.858 | 0.947 | 0.934±0.038 |
| f2 | 0.967 | 0.973 | 0.900 | 0.986 | 0.878 | 0.822 | 0.883 | 0.909 | 0.915±0.057 |
| f3 | 0.831 | 0.850 | 0.914 | 0.885 | 0.833 | 0.933 | 0.900 | 0.875 | 0.878±0.038 |
| f4 | 0.914 | 0.791 | 0.939 | 0.880 | 0.887 | 0.913 | 0.924 | 0.915 | 0.895±0.046 |
| f5 | 0.892 | 0.791 | 0.937 | 0.866 | 0.883 | 0.903 | 0.917 | 0.913 | 0.888±0.045 |
| f6 | 0.906 | 0.837 | 0.908 | 0.905 | 0.867 | 0.930 | 0.936 | 0.929 | 0.902±0.034 |
| f7 | 0.663 | 0.875 | 0.803 | 0.732 | 0.892 | 0.728 | 0.883 | 0.917 | 0.812±0.094 |
| f8 | 0.889 | 0.778 | 0.852 | 0.912 | 0.874 | 0.905 | 0.759 | 0.901 | 0.859±0.059 |
| f9 | 0.796 | 0.904 | 0.909 | 0.825 | 0.911 | 0.864 | 0.929 | 0.922 | 0.883±0.049 |
| f10 | 0.871 | 0.808 | 0.960 | 0.838 | 0.783 | 0.906 | 0.821 | 0.898 | 0.861±0.059 |

Table II: Correlation values between predicted and label images for 10-fold cross validation.

| Case | 1 | 2 | 3 | 4 | 5 | 6 | 7 | 8 | Average±S.D. |
|---|---|---|---|---|---|---|---|---|---|
| f1 | 0.881 | 0.849 | 0.891 | 0.880 | 0.959 | 0.939 | 0.916 | 0.949 | 0.908±0.039 |
| f2 | 0.850 | 0.763 | 0.912 | 0.702 | 0.927 | 0.910 | 0.942 | 0.918 | 0.866±0.088 |
| f3 | 0.847 | 0.915 | 0.901 | 0.945 | 0.892 | 0.934 | 0.938 | 0.888 | 0.908±0.033 |
| f4 | 0.695 | 0.805 | 0.729 | 0.853 | 0.918 | 0.930 | 0.955 | 0.872 | 0.845±0.095 |
| f5 | 0.696 | 0.797 | 0.682 | 0.854 | 0.915 | 0.910 | 0.951 | 0.876 | 0.835±0.101 |
| f6 | 0.836 | 0.807 | 0.846 | 0.839 | 0.832 | 0.920 | 0.948 | 0.958 | 0.873±0.059 |
| f7 | 0.758 | 0.891 | 0.830 | 0.840 | 0.910 | 0.832 | 0.826 | 0.948 | 0.854±0.059 |
| f8 | 0.899 | 0.869 | 0.792 | 0.908 | 0.896 | 0.934 | 0.863 | 0.890 | 0.881±0.042 |
| f9 | 0.888 | 0.873 | 0.912 | 0.918 | 0.916 | 0.880 | 0.923 | 0.910 | 0.903±0.019 |
| f10 | 0.918 | 0.885 | 0.962 | 0.863 | 0.864 | 0.925 | 0.891 | 0.931 | 0.905±0.035 |

## IV. Discussions and Conclusion

In this work, lung motion and HU changes between the end of inhalation and end of exhalation phases is explored by a CNN model to generate ventilation images. Although the explicit model format is unknown, CNN provides the parameters and network structures to predict ventilation images using end-of-inspiration and end-of-expiration CT images as input. This model integrates image motion, morphology, and HU changes. The proposed method does not involve the calculation of DIR, which helps reduce the uncertainty associated with the DIR algorithms.

In this preliminary study, labeled images are the ventilation images calculated using the DIR and HU change methods because of the limitation to access direct functional ventilation images. Any error in the labeled imaged will affect the training results. In future work, we will evaluate other lung function imaging modalities such as SPECT, MRI with hyperpolarized He-3, or nuclear medicine ventilation-perfusion images as labeled data for the training. The expanded validation will help demonstrate the potential robustness of the CNN model approach to calculating ventilation images.

The input data used in this study only contains two phases as the labeled images were generated from the same paired images. The proposed CNN-based method can be modified to incorporate other phases as input, especially when label data are the direct ventilation images. The performance of CNN-based method could be further improved when the input contains images from more phases by designing an appropriate architecture.

In summary, we have demonstrated the feasibility of using deep CNN to generate ventilation images from 4DCT. The predicted ventilation images after training show a high degree of similarity to the label data. With direct ventilation images as label data and more images from various phases as input, the method is expected to generate better prediction images.

## Acknowledgements

This work was partially supported by grant NIH R01CA200817 (YV, EC, RC, TG) and NIH R01 EB020366 (YZ, LC, JW). We would thank Dr. Damiana Chiavolini for helping with the editing.